\documentclass[twocolumn]{aastex6}
\usepackage{graphics}
\usepackage{grffile}
\usepackage{rotating}
\usepackage{amsmath}            

\usepackage{url}

\usepackage{color}
\usepackage{times}
\usepackage[normalem]{ulem}             
\usepackage{cancel}  
\usepackage{verbatim}				
\usepackage{mathrsfs}
\usepackage[mathscr]{euscript}

\DeclareMathAlphabet{\mathpzc}{OT1}{pzc}{m}{it}

\usepackage{tablefootnote}

\newcommand{\be}{\begin{eqnarray}}
\newcommand{\ee}{\end{eqnarray}}

\newcommand{\EM}{{\rm EM}}
\newcommand{\DM}{{\rm DM}}

\newcommand{\RM}{{\rm RM}}

\newcommand{\aau}{a_{\rm AU}}
\newcommand{\df}{d_{\rm f}}
\newcommand{\nuf}{\nu_{\rm f}}

\newcommand{\up}{u^{\prime}}

\newcommand{\xs}{x_{\rm s}}
\newcommand{\us}{u_{\rm s}}

\newcommand{\xobs}{x_{\rm obs}}
\newcommand{\uobs}{u_{\rm obs}}

\newcommand{\rF}{r_{\rm F}}
\newcommand{\uF}{u_{\rm F}}

\newcommand{\ubar}{\overline{u}}

\newcommand{\re}{r_{\rm e}}

\newcommand{\DMunits}{{\rm pc~cm$^{-3}$}}

\newcommand{\alphamin}{\alpha_{\rm min}}
\newcommand{\Gg}{G}
\newcommand{\Gp}{G_{\rm p}}

\newcommand{\dsl}{d_{\rm sl}}
\newcommand{\dlo}{d_{\rm lo}}
\newcommand{\dso}{d_{\rm so}}

\newcommand{\dslkpc}{d_{\rm sl, {kpc}}}

\newcommand{\DMlens}{\DM_{\ell}}

\newcommand{\tg}{t_{\rm g}}
\newcommand{\tdm}{t_{\rm DM}}

\newcommand{\tgzero}{t_{\rm g_0}}
\newcommand{\tdmzero}{t_{\rm DM_0}}

\newcommand{\mus}{\mu {\rm s}}		

\newcommand{\xvec}{{\mathbf{x}}}

\newif\ifnotes

\notestrue
\notesfalse

\begin{document}

\title{
Lensing of Fast Radio Bursts by  Plasma Structures in Host Galaxies
}
\shorttitle{Plasma Lensing of FRBs}
\shortauthors{Cordes et al. }


\email{cordes@astro.cornell.edu}

\author{J.~M.~Cordes}
\affiliation{Cornell Center for Astrophysics and Planetary Science and Department of Astronomy, Cornell University, Ithaca, NY 14853, USA}

\author{I. Wasserman}
\affiliation{Cornell Center for Astrophysics and Planetary Science and Department of Astronomy, Cornell University, Ithaca, NY 14853, USA}

\author{J.~W.~T.~Hessels}
\affiliation{ASTRON, Netherlands Institute for Radio Astronomy, Postbus 2, 7990 AA, Dwingeloo, The Netherlands}
\affiliation{Anton Pannekoek Institute for Astronomy, University of Amsterdam, Science Park 904, 1098 XH Amsterdam, The Netherlands}

\author{T.~J.~W.~Lazio}
\affiliation{Jet Propulsion Laboratory, California Institute of Technology, Pasadena, CA 91109, USA}

\author{S.~Chatterjee}
\affiliation{Cornell Center for Astrophysics and Planetary Science and Department of Astronomy, Cornell University, Ithaca, NY 14853, USA}

\author{R.~S.~Wharton}
\affiliation{Cornell Center for Astrophysics and Planetary Science and Department of Astronomy, Cornell University, Ithaca, NY 14853, USA}


\begin{abstract}
Plasma lenses in the host galaxies of fast radio bursts (FRBs) can strongly modulate  FRB amplitudes  for a wide range of distances, including the  $\sim $ Gpc  distance of the repeater FRB121102.   To produce caustics, the  lens'
dispersion-measure depth ($\DMlens$), scale size ($a$), and distance from the source ($\dsl$) must satisfy
$\DMlens \dsl / a^2 \gtrsim  0.65~ {\rm pc^2 \ AU^{-2} \ cm^{-3}}$.   
Caustics  produce strong magnifications ($\lesssim 10^2$) on short time scales ($\sim$ hours to days and perhaps shorter) along with narrow, epoch dependent spectral peaks (0.1 to 1~GHz).  However, strong suppression also occurs  in  long-duration ($\sim$ months) troughs.  
When bursts are multiply imaged, they 
will arrive differentially by $< 1~\mu$s to tens of ms  and  they will show different apparent dispersion measures,  $\delta\DM_{\rm apparent} \sim 1$~pc~cm$^{-3}$.   Arrival time perturbations  may mask any underlying periodicity with
period $\lesssim 1$~s.   
When arrival times  differ by less than the burst width, interference effects in dynamic spectra will be seen.   
Strong lensing requires source sizes smaller than $({\rm Fresnel~scale)^2} / a$,
 which can be satisfied by compact objects
such as neutron star magnetospheres  but not by AGNs.
Much of the phenomenology
of the repeating fast radio burst source FRB121102 can be accounted for with such lenses. 
The overall picture can be tested by obtaining  
wideband spectra of bursts (from $<1$ to 10~GHz  and possibly higher), which can also be used to 
 characterize the plasma environment near FRB sources.   
 A rich variety of phenomena is expected from  an ensemble of  lenses near the FRB source.    We discuss constraints on densities, magnetic fields, and locations of plasma lenses related to requirements for lensing to occur. 
\end{abstract}



\section{Introduction}

Fast radio bursts (FRBs) are millisecond duration pulses that show dispersive arrival times  like those seen from Galactic pulsars and that are consistent with the cold plasma dispersion law \citep[e.g.][]{tsb+13}.   The  dispersion measures (DM, the column  density of free electrons along the line of sight) of FRBs are too large to be accounted for by Galactic models of the ionized interstellar medium (ISM)
in the Milky Way \citep[e.g.][]{cl02, 2016arXiv161009448Y}.  However it is only with the measurement of the redshift of the host galaxy  firmly associated with the repeating FRB121102 \citep[][]{2017arXiv170101098C, 2017ApJ...834L...8M, 2017ApJ...834L...7T} that the extragalactic nature of an FRB has been confirmed without qualification.   It is likely that most if not all of the other $\sim 20$ reported FRBs are also extragalactic, though it is unclear whether FRB121102 is representative  of all FRBs in other respects. 

The multiple bursts obtained from FRB121102 display striking spectral diversity within the passbands of hundreds of MHz used at $\nu\sim 1.5$~GHz.   Single burst spectra show rises toward lower frequencies in some cases and rises toward higher frequencies  or midband maxima in other; burst shapes also show evolution with frequency across the band \citep[][]{sch+14, ssh+16a, ssh+16b}.   Very surprising are the changes in spectral signatures that occur on time scales less than one minute.  The repetition rate of the source is episodic and intermittent, with quiet periods of weeks to months separating observation epochs where bursts occur.   A detailed study of burst occurrence rates and  morphologies   in time and frequency is forthcoming (J. Hessels, in preparation). 


These phenomena suggest that  extrinsic propagation phenomena may play a role along with spectro-temporal variability intrinsic to the source.    Gravitational lensing or microlensing by themselves cannot be responsible for burst magnification because burst amplitudes of FRB121102 are highly chromatic.    Standard interstellar scintillation \citep[][]{1990ARA&A..28..561R}  from the Milky Way can also be ruled out because diffractive scintillations (DISS) are highly quenched by bandwidth averaging for the line of sight through the Galactic plane to FRB121102.   
Many  `scintles'  with characteristic bandwidths $\sim 60$~kHz at 1.4~GHz 
\citep[estimated with the NE2001 model; ][]{cl02}
  are averaged over the  300 to 600~MHz bandwidths used at Arecibo,  yielding a DISS modulation fraction of only a few per cent. Refractive scintillations typically only cause modulations of tens of per cent and vary on time scales significantly longer
  than DISS; also, they do not introduce any narrow spectral structure.
  Strong focusing by plasma lenses is the most likely process because it can produce large variations (factors of ten to 100) that could account for much of the intermittency seen  from FRB 121102.

Plasma lensing from $\sim$~AU structures in the Milky Way are consistent with `extreme scattering'  events (ESEs) seen in the light curves of a few active galactic nuclei  
\citep[AGNs;][]{1987Natur.326..675F, 2016Sci...351..354B} and pulsars along with timing perturbations for the latter objects \citep[][]{2015ApJ...808..113C}.  
Required densities indicate they are over-pressured with respect to the mean ISM but by an amount that is minimized if events occur when plasma filaments or sheets are edge on to the line of sight.   Very strong overpressure in  about 0.05\% of neutral gas in the cold ISM \citep[][]{2011ApJ...734...65J} is attributed to turbulence driven by stellar winds or supernovae.  There is no consensus on the relationship of  ESE lenses to other interstellar structures   \citep[e.g.][]{1987Natur.328..324R, 1998ApJ...498L.125W, 2014MNRAS.442.3338P}. 
However, a common feature may be edge-on alignment of structures inferred for both Galactic plasma lenses and cold HI 
\citep[][]{1987Natur.328..324R, 1997ApJ...481..193H}.   It is therefore especially notable that 
 plasma lensing is    manifested in distinctive DM variations of the Crab pulsar  from dense filaments in the Crab Nebula. \citet[][]{2011MNRAS.410..499G} infer diameters $\sim 2$~AU
and DM depths $\sim 0.1$~\DMunits\ from DM variations  whereas observations of optical  filaments  indiciate larger sizes $\sim 1000\, {\rm AU} \times 0.5$~pc with electron densities $n_{\rm e} \sim 10^4$~cm$^{-3}$ \citep[][]{1985ARA&A..23..119D}.

In this paper we consider lensing from one-dimensional plasma structures in the host galaxy of FRB121102, like the Gaussian lens  analyzed by \citet[][]{1998ApJ...496..253C}.     One-dimensional lenses show complex properties that may be sufficient to account for the observations of the repeating FRB.  Moreover, one-dimensional structures may be physically relevant given that long  filaments  are seen in the Crab Nebula and it is possible   that FRB121102 is associated with a young neutron star in a supernova remnant \citep[e.g.][]{pc15, cw16, 2016MNRAS.458L..19C, 2017arXiv170102370M, 2017arXiv170101098C, 2017ApJ...834L...8M, 2017ApJ...834L...7T}.

In \S~\ref{sec:optics} we derive the general properties of plasma lenses with emphasis on those that are much closer to the source than to us.     
In \S~\ref{sec:examples} we discuss their application to FRBs while 
\S~\ref{sec:crab}  discusses filaments in the Crab Nebula and their lensing properties.  
 \S~\ref{sec:hosts} presents detailed results for an example lens that can account for many of the properties of FRB121102.   We discuss our results and make conclusions in \S~\ref{sec:summary}. 
Appendix \ref{app:KDI} elaborates on our use of the Kirchhoff diffraction integral (KDI) of the Gaussian lens.

\section{Optics of a Single Plasma Lens}
\label{sec:optics}

\citet[][]{1998ApJ...496..253C} analyzed the properties of a one-dimensional (1D) Gaussian plasma lens as a means for understanding discrete events in the light curves of AGNs.   Their analysis assumed incidence of plane waves on Galactic lenses of the form
$\DM(x) = \DMlens \exp{(-x^2/a^2)}$, that  yield a phase perturbation $\phi_\lambda = -\lambda \re \DM(x)$, where  $\lambda$ is  the radio wavelength and  $\re$ is  the classical electron radius. The lens has a characteristic scale $a$ and the maximum electron column density through the lens is $\DMlens$.  
 While a positive column density $\DMlens > 0$ acts as a diverging lens,  rays that pass through different parts of the lens can converge to produce caustics.  Two dimensional lenses and the inversion of measurements of lensing events into DM profiles have been considered  by \citet[][]{2016ApJ...817..176T}. 

We extend the analysis of \citet[][]{1998ApJ...496..253C}  to the more general case where a source is at a finite distance from the lens.  
Let $\dsl$ and $\dso$
be the distances from the source to the lens and observer, respectively, which define the lens-observer distance
$\dlo = \dso-\dsl$.   

Transverse coordinates    in the source,  lens, and observer's planes are $\xs, x, $ and $\xobs$, respectively, and
we define dimensionless coordinates  $\us = \xs/a$, $u = x/a$, and $\uobs = \xobs/a$ by scaling them by the lens scale $a$. 
The lens is centered on $u=0$.  We also define  a combined offset 
\be
\up = (\dlo/\dso) \us + (\dsl/\dso) \uobs.
\label{eq:up}
\ee  
Offsets or motions of the source dominate $\up$ for lenses close to the source ($\dsl/\dso \ll 1$),
while the observer's location dominates $\up$ for Galactic lenses with  $1-\dsl/\dso \ll 1$.

The lens equation in geometric optics corresponds to  stationary phase solutions for $u$ of  the Kirchhoff diffraction integral (Appendix~\ref{app:KDI}), 
\be
u(1+\alpha e^{-u^2}) = \up,
\label{eq:lens_eq}
\ee
where $\alpha$ is  the dimensionless parameter 
\be
\alpha &=& \frac{\lambda^2 \re \DMlens}{\pi a^2}  \left(\frac{\dsl \dlo}{\dso}\right)
=  \frac{ 3430\, \DMlens \dsl}{(\nu \aau)^{2} }  \left( \frac{\dlo}{\dso}\right),
\label{eq:alpha}
\ee
where  $\nu$ is in GHz,  $\DMlens$ has standard units of \DMunits, the lens scale $a$ is  in AU, and the source-lens distance $\dsl$ is in kpc.    Ray tracings depend only on $\alpha$ and therefore are the same for a wide variety of lens sizes, DMs, and locations. 

Electron density {\it enhancements} give positive values of $\alpha$ while voids in an otherwise uniform density correspond to $\alpha < 0$ and yield a converging rather than a diverging lens.   We restrict our analysis to density enhancements in this paper  because they are sufficiently rich in phenomena to perhaps account for FRB properties.  However, we  note that density voids also may be relevant to plasmas surrounding FRB sources.   Most of the equations presented here are valid for 
$\DMlens < 0$, but require  an absolute value in some cases.  Lensing is qualitatively similar to the
$\DMlens > 0$ case but the locations of caustics differ. 

While we use the 1D Gaussian lens in most of this paper, the basic approach can be applied to an 
arbitrary 2D perturbation, $\DM(\xvec) = \DMlens \psi(\xvec)$ where $\psi(\xvec)$ is a dimensionless function with unity maximum.  In this case the lens equation is
${\bf u}^{\prime} = {\bf u} +  \alpha a \nabla_{\bf u} \psi({\bf u})$ and $a$ is  a characteristic scale of $\psi({\bf u})$. 

\begin{figure}[t!]
\begin{center}
\includegraphics[scale=0.42]
{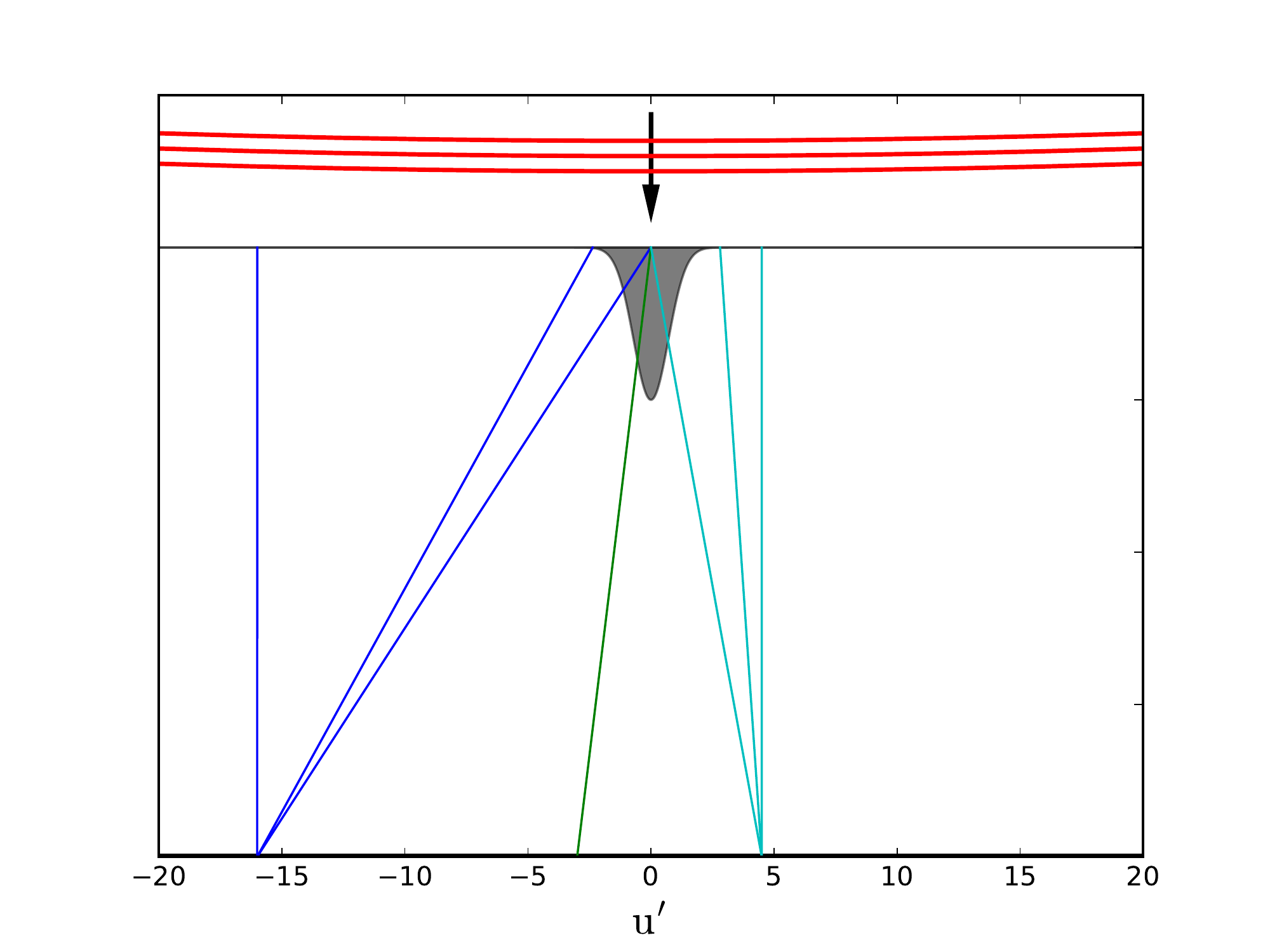}
\caption{Ray traces  from the lens plane to three locations in the transverse plane.    Three rays reach the two transverse positions at $\up \sim -16$ and $+4.5$ while only one reaches $\up \sim -3.5$.   The ray tracing is for   $\alpha=1524$ which corresponds to $\DMlens = 1$~\DMunits\, $a = 1$~AU, $\dsl = 1$~kpc, $\dso = 1$~Gpc, and $\nu = 1.5$~GHz but could  correspond  to other combinations of parameters.    For lenses near a fixed source at $\us = 0$,  the change in observer's position $\uobs$ needed to yield a unit change in $\up$ is  $\dso/\dsl \gg 1$. 
\label{fig:raytraces}
}
\end{center}
\end{figure}

For a given offset $\up$ there are  either one or three solutions for $u$, as shown in Figure~\ref{fig:raytraces} for three values of
 $\up$.     For this figure $\alpha = 1524\nu^{-2}$, which is given  by
  $\DMlens = 1$~pc~cm$^{-3}$, $a=1$~AU, $\dso = 1$~Gpc and $\dsl = 1$~kpc but could also correspond to substantially different lens parameters, such as a much smaller $\dsl$ combined with larger $\DMlens/a^2$.   In this case $\up$ is totally dominated by motions of the source because $\dsl/\dso = 10^{-6}$.   However, the same ray traces apply to the Galactic case when the lens  is 1~kpc from the Earth.  In that case, motions of the observer and lens will dominate changes in the line of sight. 
  
 \begin{figure}[h!]
\begin{center}
\includegraphics[scale=0.42]
{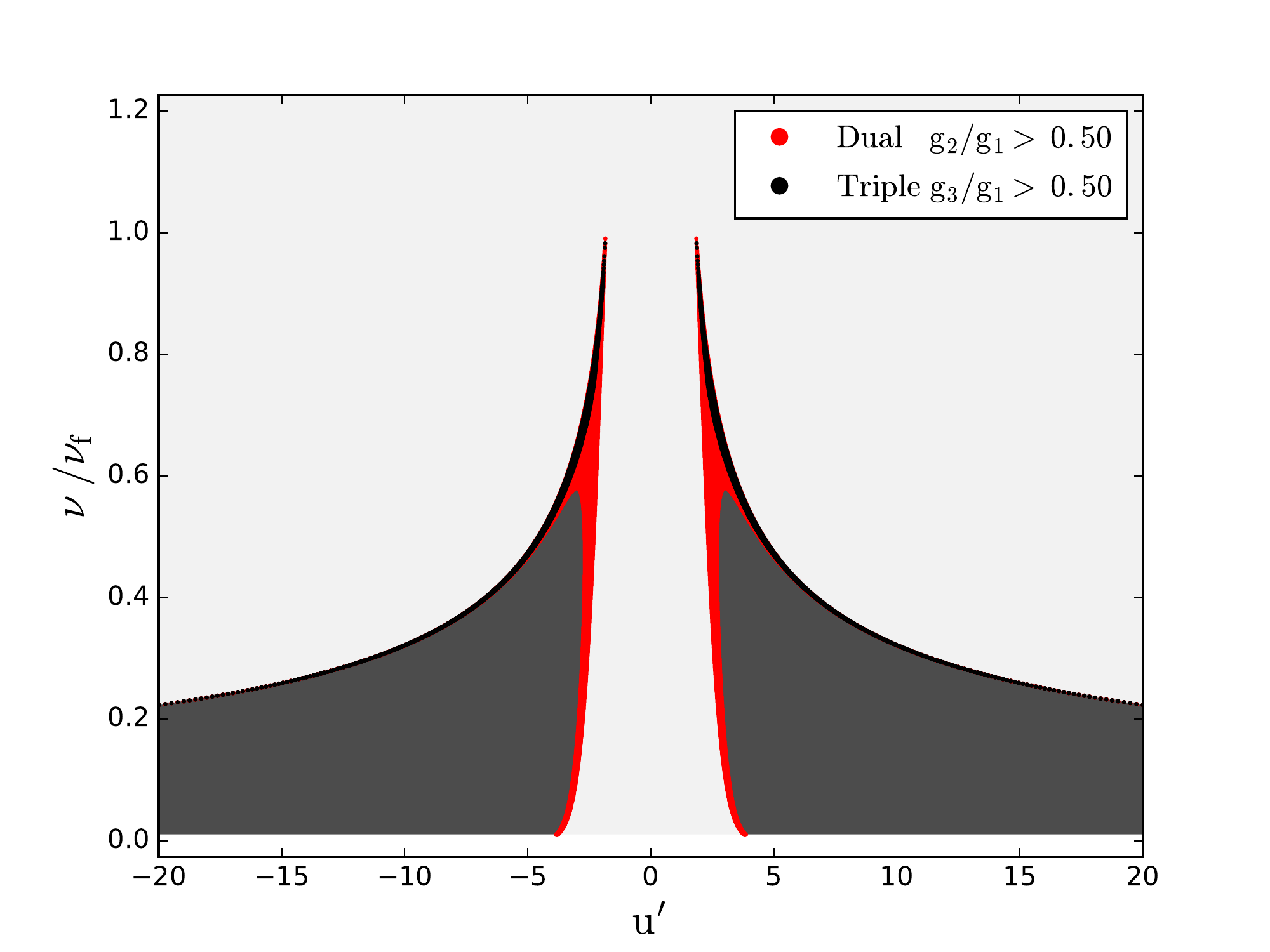}
\caption{
Regions in the frequency -- location plane where either one or three images are seen. 
The frequency $\nu$  has been normalized by the focal frequency $\nuf$ (Eq.~\ref{eq:nuf}) and the location $\up$ is a linear combination of the source's and observer's positions given by Eq.~\ref{eq:up}. 
Off white indicates the region where single-images occur while three images occur elsewhere.
Red and black show  where two or three images occur with approximately equal gain ratios. 
The red region is where the second largest gain is at least 50\% of the largest gain while black indicates where
all three images have gains within 50\% of each other. 
\label{fig:nimages}
}
\end{center}
\end{figure}

Figure~\ref{fig:nimages} shows the regions  in the $\nu$-$\up$ plane where either one or three subimages are seen.  The frequency
axis is expressed in units of a focal frequency $\nuf$ defined below.   In dimensional units triple images are seen up to nearly 40~GHz
for the case shown.    For smaller values of $\alpha$, region of  triple images moves to lower
frequencies. 
A single burst from an FRB source can therefore  be manifested as three sub-bursts with different amplitudes, DMs, and arrival times for some observer locations and frequencies.  

\subsection{Amplification}

In the geometrical optics regime, the  focusing  or defocusing of  incident wavefronts yields a 
 `gain' (or amplification) given by the  stationary-phase solution for  $u$  \citep[Appendix~\ref{app:KDI};][]{1998ApJ...496..253C},
\be
\Gg = \left\vert 1 + \alpha(1-2u^2)  e^{-u^2}\right\vert^{-1}.
\label{eq:gain}
\ee

\begin{figure}[t!]

\begin{center}
\includegraphics[scale=0.42]
{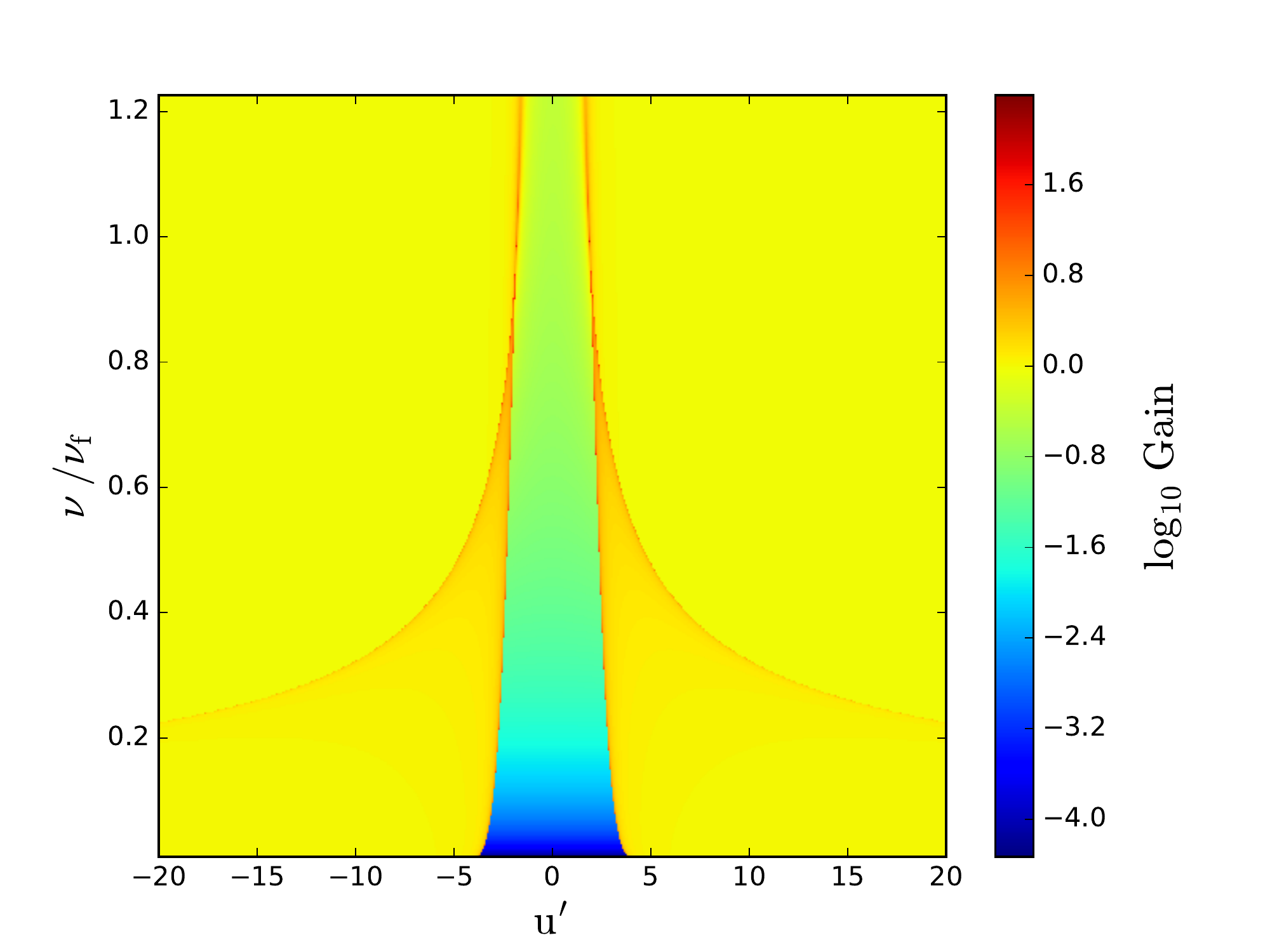}
\caption{Total gain summed over all rays that reach an observer vs. frequency (normalized by the focal frequency $\nuf$) and transverse location. 
\label{fig:gain}
}
\end{center}
\end{figure}

\begin{figure}[t!]
\begin{center}
\includegraphics[scale=0.42]
{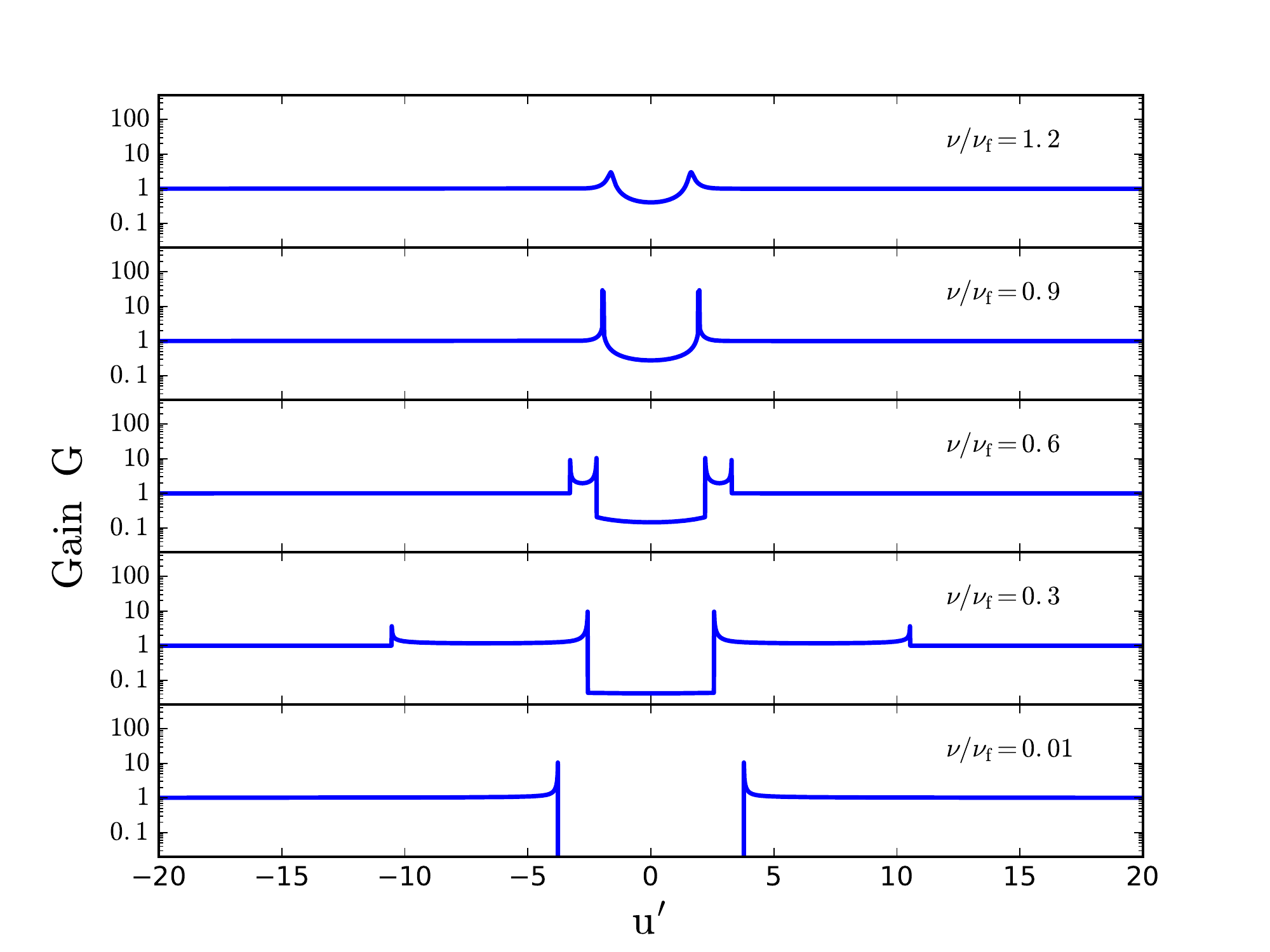}
\caption{Spatial slices of the gain $G$ for a few frequencies (normalized by the focal frequency $\nuf$). 
\label{fig:gainslices}
}
\end{center}
\end{figure}

The variation of $\Gg$ with frequency and transverse location in  Figure~\ref{fig:gain} shows that  the highest values occur where there are three images and the lowest values are in a central trough that is aligned with the direction to  the lens. 
Horizontal slices  through this plane are shown in Figure~\ref{fig:gainslices} at a few frequencies.
For triplet images, the total gain is the sum of the gains for the three screen positions $u$ that contribute.

The minimum gain in the trough is $G_{\rm min} = 1 / (1 + \alpha)$  at $u = \up = 0$.
Some locations are also focal points where $G\to\infty$. These occur
 for  values of $\alpha$,
\be
\alpha_\infty = \frac{e^{u^2}}{2u^2 -1},
\label{eq:alphareq}
\ee
where $u=u(\alpha, \up)$ is the stationary phase solution of Eq.~\ref{eq:lens_eq}. 
No infinities occur for $\vert u \vert < 1/\sqrt{2}$ because $\Gg<1$ when $\alpha$ is restricted to positive values.

The required $\alpha_\infty$ diverges at $\vert u\vert =1/\sqrt{2}$ 
and decreases to a minimum $\alphamin = e^{3/2}/2 = 2.24$ at
$\vert u\vert =\sqrt{3/2} = 1.22$.    Diverging gains at larger $\vert u\vert > 1.22 $  require exponentially 
increasing  values of $\alpha_\infty$ corresponding to very large
densities and distances or to very small lenses that may  also produce free-free absorption or may  be physically implausible. 

To identify locations of infinities numerically, we
eliminate $\alpha e^{-u^2}$  by combining Eq.~\ref{eq:lens_eq} and \ref{eq:gain}   to obtain the algebraic equation,  
$2u^3 - 2u^2 \up + \up = 0$.
Solutions for $u$  yield the values of $\alpha_\infty$ from Eq.~\ref{eq:alphareq} needed for the apparent infinity.
Actual physical optics gains are finite and can be calculated either through exact evaluation of the Kirchhoff diffraction integral
or by integrating the third-order polynomial expansion of the total phase (Appendix~\ref{app:KDI}).   
At $\alpha = \alphamin$ and $u=\sqrt{3/2}$ the maximum gain  is 
\be
G_{\rm max} \sim a / \rF,
\ee
where 
$\rF = \sqrt{\lambda \dsl\dlo/2\pi\dso}$ is the Fresnel scale.  Larger lenses therefore give larger maximum gains. 
We note in Appendix~\ref{app:KDI} that the true gain at this location is affected by catastrophe optics \citep[][]{Berry1980257}, which describes specific fundamental forms for the large variation of the gain with small changes in position or frequency.   

The actual expected gain also depends on the physical size of the source and on any scattering from small-scale irregularities prior to encountering the lens or in the lens itself.   A full discussion of these issues is beyond the scope of the paper, although a constraint on source size is given in \S~\ref{sec:sourcesize}.

 \subsection{Focal Distance and Focal Frequency}


Bursts originating from 
lens-observer distances  $\dlo$ larger than a focal distance  $\df$  reach the observer on multiple paths and are  affected by caustics in light curves.   The requirement for multiple images,  $\alpha > \alphamin$, gives
\be
\df(\nu) &=&  \dlo \left( \frac{\alphamin}{\alpha} \right)
        =  \frac{\pi (a \nu)^2 \alphamin}{\re c^2 \DMlens}  \left(\frac{\dso}{\dsl}\right)
\nonumber \\
	&\approx& 0.65\, {\rm Mpc}\times \frac{ (\aau\nu)^2 }{ \DMlens} \left(\frac{\dso/\dsl}{10^6}\right),
\label{eq:df}
\ee
where the coefficient applies to the case where the lens is nearer the source than the observer by
a factor of $10^6$.   Galactic cases with $\dso/\dsl \sim 1$ can have sub-pc focal distances, which corresponds to 
a regime where the gain will be close to unity. 
 
A similar analysis implies that  frequencies below the  focal frequency $\nuf$ will show ray crossings,  
\be
 \nuf &=& \nu \left( \frac{\alpha}{\alphamin} \right)^{1/2} 
 =  \frac{c}{a}  \left( \frac{\re  \DMlens }{\pi \alphamin} \frac{\dsl\dlo}{\dso} \right)^{1/2} 
 \nonumber \\
	&\approx&
	39.1~{\rm GHz} \times \frac{\DMlens^{1/2}}{\aau} \left(\frac{\dsl\dlo/\dso}{1~\rm kpc} \right)^{1/2} .
\label{eq:nuf}
\ee
In terms of the focal distance, the focal frequency is  $\nuf = \nu \sqrt{\dlo / \df(\nu)}$.

\subsection{Constraints on Lens Parameters}

In their study of  AGN light curves, \citet[][]{1998ApJ...496..253C} considered Galactic  lenses
with $\dsl/\dso \sim 1/2$ and $\dlo\dsl/\dso \sim 1~$kpc.  These give
focal values  ($\df$ and $\nuf$) similar to the path length through the Galaxy and to  the observation frequencies (2.25 and 8.1 GHz) they considered, implying that the `extreme scattering events' they analyzed  \citep[][]{1987Natur.326..675F}
were well within the caustic regime. 

Lenses embedded in host galaxies of burst sources with $\dso/\dsl \gg 1$  yield much larger focal distances but similar
focal frequencies. 

If the observer's distance is a multiple $M$  of the focal distance 
(corresponding to $\alpha = M \alphamin$), the lens parameters satisfy
\be
\frac{ a^2}{ \DMlens}  
\! 
= 
\!
\frac{\re c^2}{\pi M \alphamin \nu^2}
\left(\frac{\dsl\dlo}{\dso} \right) 
\!
 \approx \!  \frac{1531}{M\nu^2} \left(\frac{\dsl\dlo/\dso}{1~\rm kpc} \right)\! ,
\label{eq:at_focal}
\ee
where the approximate equality is for 
$a$   in AU units, $\nu$ in GHz and $\DMlens$  in \DMunits.
Eq.~\ref{eq:at_focal} implies  that identical lens parameters can satisfy the  condition
$\dlo = M\df $ for either Galactic or extragalactic lenses.

Focal distances equal to the lens distance ($M=1$) may yield a high degree of variability if, for example, a population of lenses moves across the line of sight.    Heterogeneous lenses and geometries with different focal distances will 
produce variability with a range of amplitudes and time scales.

\subsection{ Times of Arrival (TOA)}

Burst arrival times  are chromatic due to plasma dispersion along the line of sight. 
Here we analyze only the TOA perturbations from the plasma lens, which add to delays from other media along the line of sight.  For each image produced by the lens, the burst arrival time receives a  contribution $\tg$ from the geometrical path length  of the refracted ray path and from the dispersion delay through the lens.   The geometric term is
\be
\!\!\!\!\!\!\!
\tg(x) &=& (2c)^{-1} \times
\nonumber \\
&&
\!\!\!\!\!\!\!
\left[ \frac{(x-\xs)^2}{\dsl} +  \frac{(x-\xobs)^2}{\dlo}  - \frac{(\xobs-\xs)^2}{\dso} \right],
\ee
which vanishes for an unrefracted ray path with 
$x = (\dlo \xs + \dsl\xobs)/\dso$. 
The dispersion delay  $\tdm$ through the lens,
\be
 \tdm(x) &=& (c \re /2\pi \nu^2) \DM(x).
\ee  %


Using  dimensionless coordinates and combining the two perturbations, the TOA is
\be
t(u) = \tgzero (u-\up)^2 + \tdmzero e^{-u^2},
\ee
where the geometrical delay coefficient is  
\be
\tgzero &=& \frac{a^2}{2c} \left( \frac{\dso}{\dsl\dlo} \right)
\label{eq:tgzero}
\ee
and the dispersion coefficient is
\be
\tdmzero &=& \alpha \tdmzero =  \frac{c \re \DMlens}{2\pi \nu^2} = 4.149~{\rm ms} \times \frac{\DMlens}{\nu^2},
\ee
which  is evaluated for frequencies in GHz and \DM\ in standard units of pc~cm$^{-3}$. 

For a  specific  solution to the lens equation,  $u = u(\up, \alpha)$,   the TOA has the simple form
\be
t(u) &=& 
		\tdmzero e^{-u^2} \left(1 + \alpha u^2 e^{-u^2}  \right),
\ee
where the first term is the dispersive contribution. 
Different images emanate from different positions $u$ in the lens plane, so they will have different DM values
$\DMlens e^{-u^2}$ and arrival times $t(u)$ that include a geometric contribution. 
A larger lens closer to the source increases the geometric perturbation while the dispersion delay can be substantially larger
than $\tg$  if $\alpha \gg 1$.    
Evaluating Eq.~\ref{eq:tgzero} using the focal constraint in Eq.~\ref{eq:at_focal} for a distance
$\dlo = M\df$, we obtain
\be
\tgzero = \frac{\tdmzero}{M\alphamin} \approx \frac{0.45\, \tdmzero}{M}. 
\ee
For $M=1$ the terms are comparable but their variations with location and frequency generally differ.


\subsection{Apparent Burst DMs}

\newcommand{\DMhat}{\widehat{\DM}}

For a fixed observer and source (i.e. fixed $\up$), a shift in frequency changes the position $u$ of a ray that
 alters both the geometric and dispersive TOA contributions.  Consequently, estimates of DM from multifrequency
observations will reflect both contributions and not just the true dispersion.  
Using the derivative $dt/d\nu$  to estimate \DM\ for fixed $\up$ (stationary source and observer) and
using the solution $u = u(\up, \alpha)$, we obtain
\be
\DMhat 
= -\frac{2\pi\nu^3}{c\re} \frac{dt}{d\nu}
= \DMlens e^{-u^2} \left(1 \pm 4\alpha G u^2 e^{-u^2} \right)
\ee
where the $\pm$ designates the parity as defined in Appendix~\ref{app:KDI}, which changes sign 
when $G^{-1} = 0$. The parity is positive   for
 $1+\alpha (1-2u^2) e^{-u^2} > 0$.   
 
 Estimating DMs from $dt/d\nu$ or from multifrequency fitting  will give discrepancies between  individual burst components that scale with $\DMlens$.    The situation will be especially complex when burst components overlap in the time-frequency plane.

\subsection{Frequency Structure in Burst Spectra}

\begin{figure}[t!]
\begin{center}
\includegraphics[scale=0.42]
{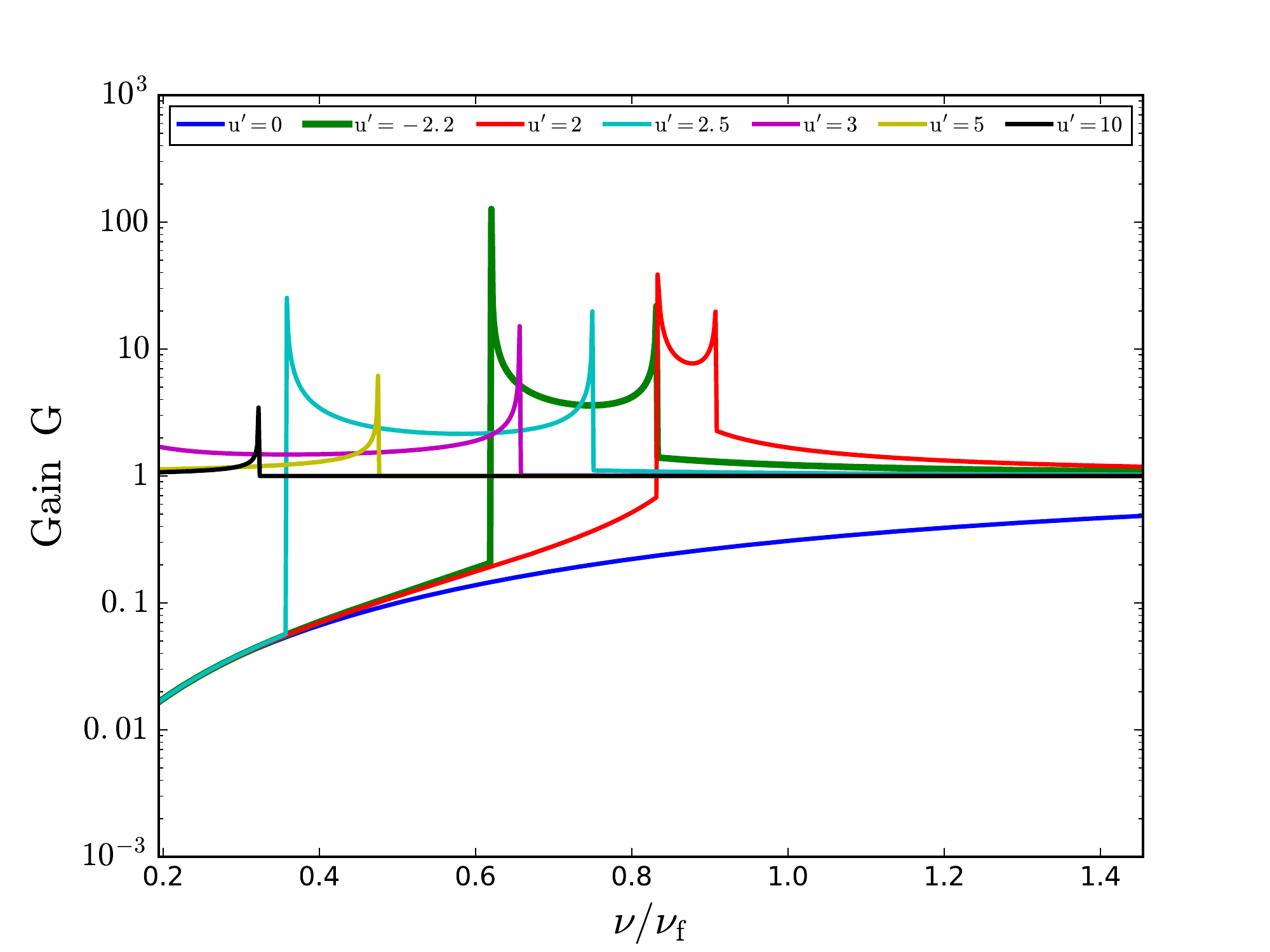}
\caption{Spectral slices of the gain $G$ at a few observer locations for $a=60$~AU and $\DM = 10$~\DMunits.
\label{fig:gainspectra}
}
\end{center}
\end{figure}

Figure~\ref{fig:gainspectra} shows gain spectra that are vertical slices of Figure~\ref{fig:gain} at various locations $\up$.  One is for the center of the gain trough at $\up = 0$.  Other slices show the generic double-cusp shape that results from traversing the three-image region in Figure~\ref{fig:nimages} and one or both single-image regions.   The spectral dependence of the gain implies that a continuum source would often be detected with significant spectral modulations that include cusp-like features (for a Gaussian lens) with widths of a few to ten percent of the observation frequency.   The shapes of gain enhancements will differ for alternative density profiles.

When a burst is multiply imaged and the extra delays in the arrival times of subimaged bursts are smaller than their widths,  the subimages will interfere, producing time-frequency structure like that seen in the dynamic spectra of pulsars.     Small lenses with $a \lesssim 1$~AU that yield sub-microsecond arrival time differences will produce frequency structure on scales of $\sim 1$ to 100~MHz.   These oscillations will modulate  the broader, caustic-induced frequency shape by amounts that depend on the gain ratio of a pair of images.   If all three images contribute significantly, the oscillation frequencies $f_{ij}$ will satisfy a closure relation, $f^{-1}_{12} + f^{-1}_{23} + f^{-1}_{31} = 0$.   Measurements of oscillations can provide a test for the presence of  multiple imaging.

\subsection{Effects of Source Size and Motion}
\label{sec:sourcesize}

A  change in the position of a point source alters the total phase $\Phi$  in the Kirchoff diffraction integral (Appendix~\ref{app:KDI}).    
Therefore, integrating over an extended source of
size $\delta\us = \delta \xs / a$ will reduce the gain if $\delta\us$ induces a phase shift $\sim \pi$.   This translates into
an upper bound on the source size, 
\be
\delta \xs \le \frac{a\pi}{\partial_{\us}\Phi} 
	&=& 
	\frac{\lambda\dsl}{2a}
	\left(\frac{2u^2-1}{u} \right)
	\left(\frac{G}{G-1} \right)
\nonumber \\
&\approx& 3\times10^9~{\rm cm}\, \frac{\dslkpc}{\aau\nu},
\ee
where the  approximate equality is for  position and gain factors of order unity, an AU-size lens, and a source-lens
distance $\sim 1$~kpc.   Higher frequencies, larger lenses,  and lenses nearer the source yield more stringent requirements on $\delta\xs$. 

 The nominal upper bound on $\delta \xs$ is similar to that required to see interstellar scintillations of Galactic pulsars
and   is easily satisfied by emission regions with transverse sizes of  a light millisecond  ($\sim 3\times10^7$~cm). 
The light cylinder of a rotating object with period $P\lesssim 1$~s has radius $\sim 5\times10^9P$~cm, so even slowly spinning objects with small emission regions inside their light cylinders will show lensing. 
Emission from larger objects, such as AGN jets will strongly attenuate the lensing  unless there is  coherent emission  from very small substructures.    

Relativistic beaming of radiation from the source could in principle reduce the illuminated portion of the lens,   but  the beam exceeds  the lens size for $\gamma < 2\times10^8\, \dslkpc / a_{\rm AU}$.    This is well satisfied, for example, by relativistic flows in the magnetospheres of neutron stars for which radio emission is estimated to be from particles 
with $\gamma \lesssim 10^3$.


\newcommand{\veffperp}{v_{\perp}}
\newcommand{\vsperp}{v_{\rm s, \perp}}
\newcommand{\vobsperp}{v_{\rm obs, \perp}}

\newcommand{\tG}{t_{\rm G}}
\newcommand{\tcaustic}{t_{\rm caustic}}
\newcommand{\ttrough}{t_{\rm trough}}

To calculate the time scale for changes in gain through caustics, we calculate
 the change in position on the screen $\delta u$ of a ray   by  taking the derivative of the lens equation Eq.~\ref{eq:lens_eq} and employing other quantities, 
\be
\delta u = G\delta \up = G\left[(\dlo/\dso) \delta\us + (\dsl/\dso) \delta\uobs \right].
\ee
The corresponding change in gain is $\delta G \approx (dG/du) \delta u$, or
\be
\frac{\delta G}{G} = \frac{4 G(G-1) u (u^2-3/2) \delta \up}{\vert 1 - 2u^2\vert} .
\ee
Motions of source and observer
combine into an effective transverse velocity,
\be
\veffperp = (\dlo/\dso) \vsperp + (\dsl/\dso) \vobsperp.
\ee
Using $\delta\up = \veffperp \tcaustic / a$ and
 $\veffperp = 100\,v_{100} \, {\rm km~s}\, $,  
the characteristic  time scale of a caustic crossing is
\be
\!\!\!\!\!\!\!
\tcaustic \sim \frac{ a(\delta G/G)}{\veffperp G^2} \left( \frac{\dso}{\dlo}\right) 
\sim
\frac{ 4.2\ {\rm hr } \times a_{\rm AU} (\delta G/G) }{v_{100}(G/10)^2 }\left( \frac{\dso}{\dlo}\right).
\nonumber\\
\label{eq:tcaustic}
\ee
The brightest caustics can yield $G > 10$ and  even smaller characteristic times. 

The gain trough centered on $\up = 0$ has a width $\Delta\up \sim 5$ (Figure~\ref{fig:gainslices}), corresponding to a
radio-dim time span, 
\be
\ttrough \sim \frac{a \Delta\up}{\veffperp} \approx 87 \,{\rm d} \left(\frac{\aau}{v_{100}}\right). 
\label{eq:trough}
\ee

\section{Single Lens Examples Relevant to FRBs}
\label{sec:examples}

Our view is that multiple plasma lenses in the host galaxy of an FRB have a variety of dispersion depths, sizes, and distances from the source.   Individually these will produce lensing effects that span a wide range of time scales, gains, and timing perturbations as described in the previous section.  However,  caustics may also result from the collective effects of two or more lenses and  the duty cycle for lensing may be larger than implied for a single lens.     A detailed analysis of multiple lenses is beyond the scope of the present paper. 

For specificity, we  consider a lens with a dispersion depth $\DMlens = 10$~\DMunits\ and  scale size $a = 60$~AU. 
These parameters, along with  the same distance parameters as before ($\dso = 1$~Gpc and $\dsl = 1$~kpc),
yield observables that are similar, at least qualitatively,  to those seen for the repeating FRB121102.   However, we emphasize that the observations do not allow any unique determination of parameter values and our choices here are simply for the purpose of illustration.   A detailed comparison with the multiple bursts detected  from the repeater FRB121102 will appear elsewhere. 

 The lens parameters give  $\alpha = 9.5\,\nu^{-2}$ and 
a focal frequency (Eq.~\ref{eq:nuf}) $\nuf \sim 2$~GHz  comparable to typical observation frequencies  used in FRB surveys.  

\begin{figure}[t!]
\begin{center}
\includegraphics[scale=0.42]
{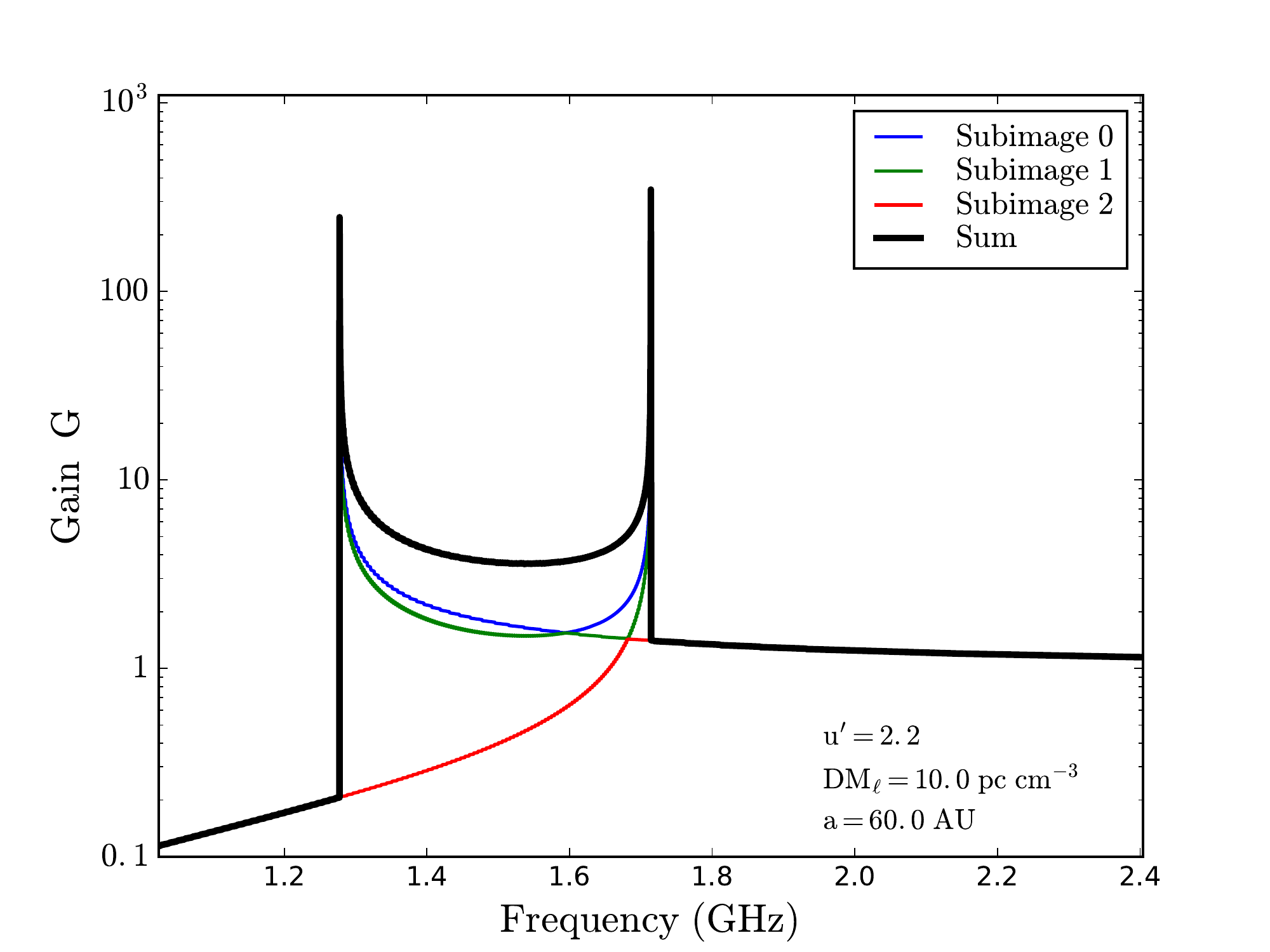}
\caption{
Gain vs. frequency for individual subimages and the summed gain  at a selected transverse position $\up=2.2 $ and a lens with
$\DMlens = 10$~\DMunits\ and $a = 60$~AU.
 \label{fig:gain_vs_frequency_10_60}
}
\end{center}
\end{figure}

Figure \ref{fig:gain_vs_frequency_10_60}  shows the gains of individual subimages along with the total  gain as a function of frequency for a particular location $\up = 2.2$.   Gain enhancements are confined  roughly  to the frequency band of 1.28 to 1.78~GHz and appear as two caustic peaks, one rising at lower frequencies, the other rising at higher frequencies.  The gain is suppressed below unity below the low-frequency caustic while it asymptotes to unity above the high-frequency caustic.  

Observations of FRB121102 have spanned $\sim 4$~yr to date and would correspond to a variation in $\up$ that depends on
the geometry and relevant transverse velocities, which are dominated by source  or lens motions for lenses in host galaxies.  

\begin{figure*}[t!]
\begin{center}
\includegraphics[scale=0.42]
{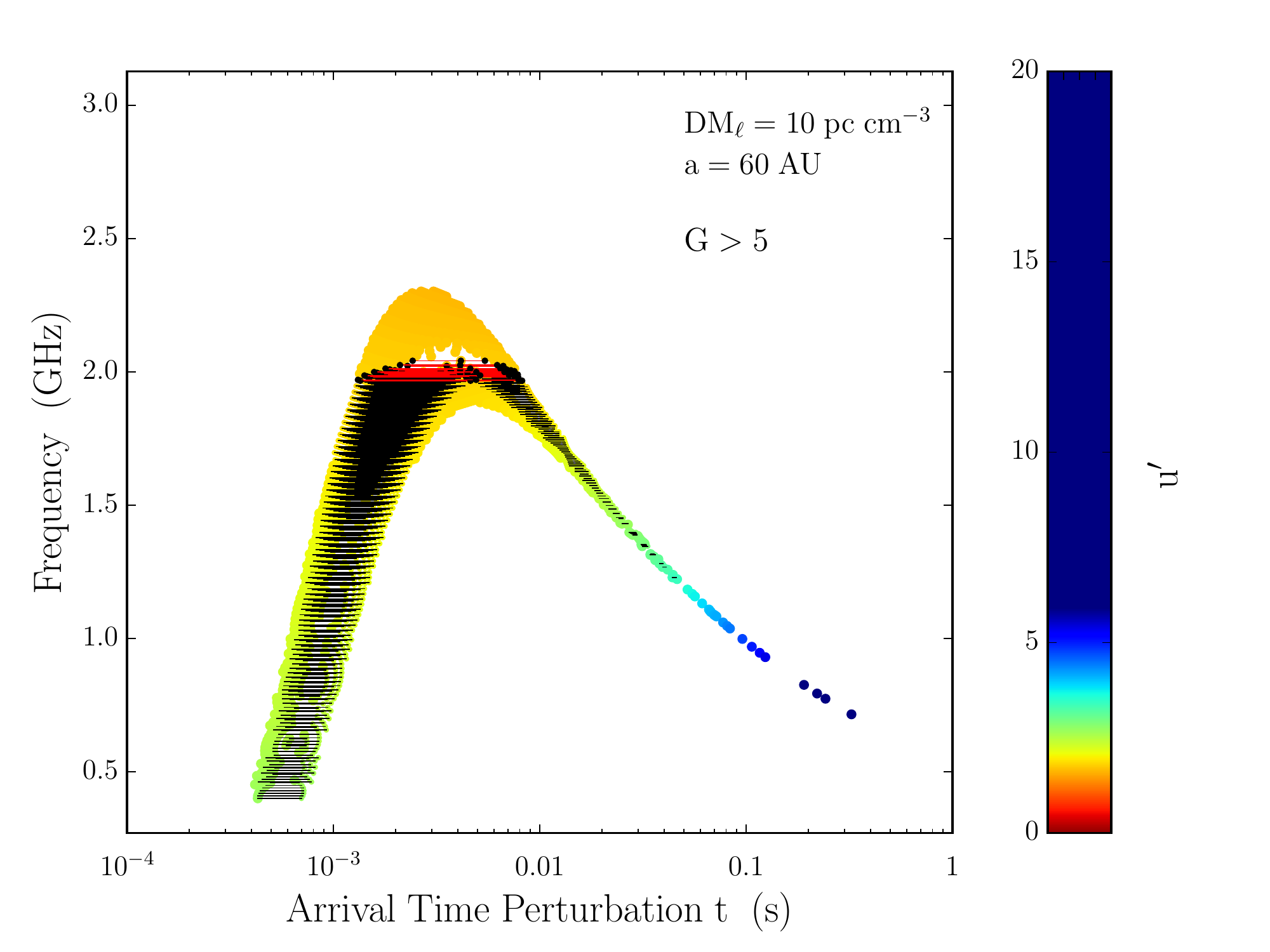}
\includegraphics[scale=0.42]
{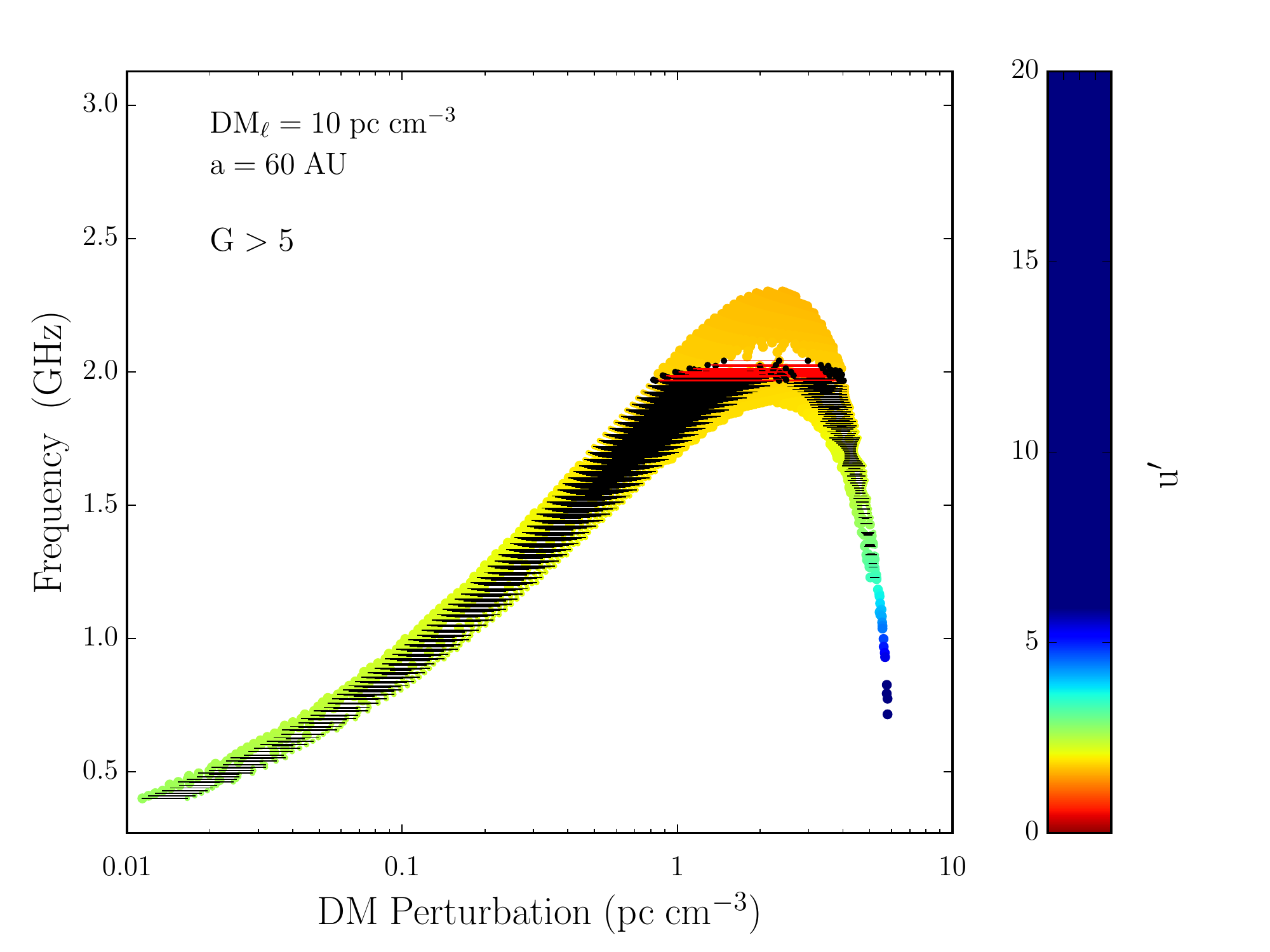}
\includegraphics[scale=0.42]
{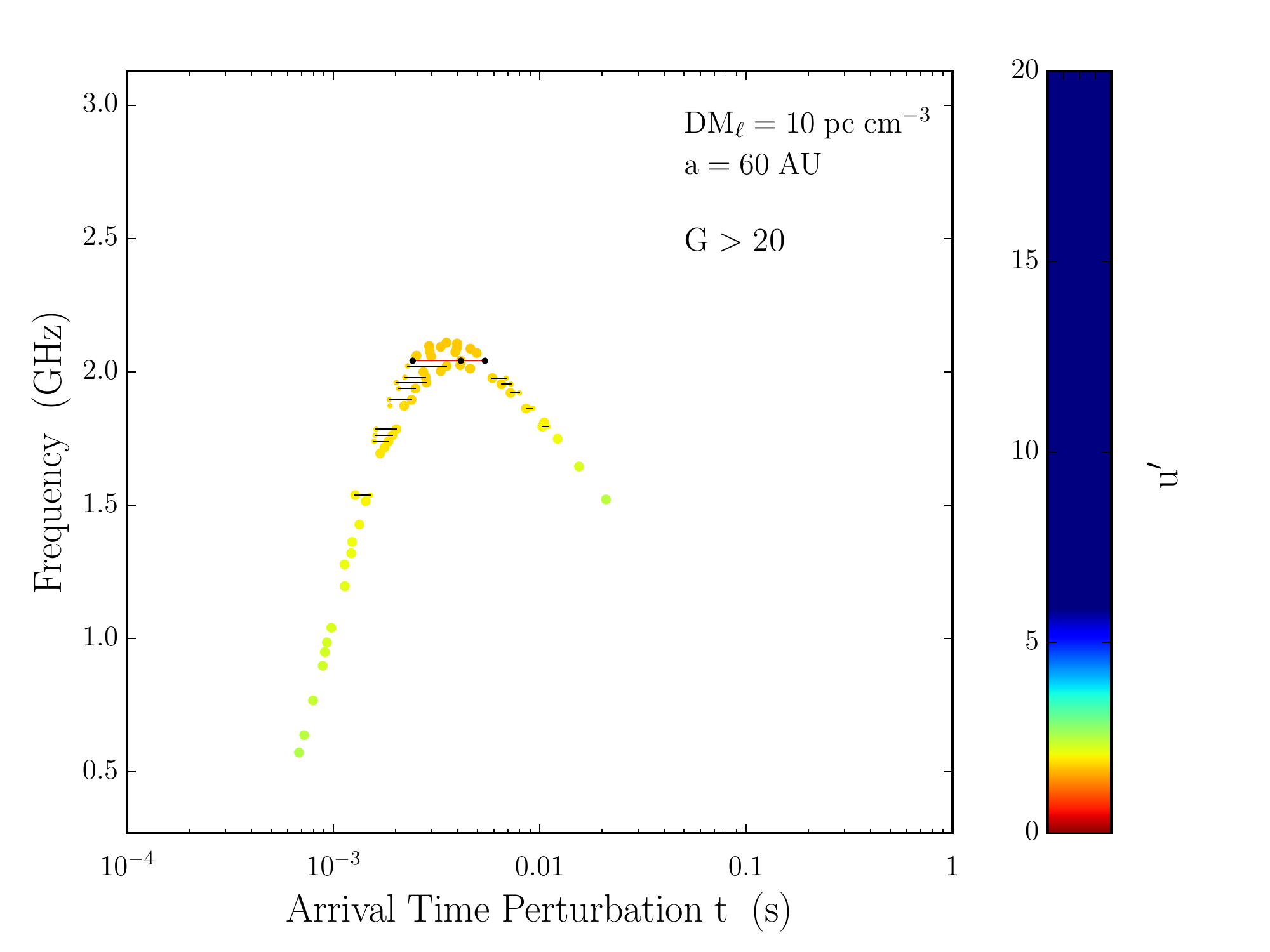}
\includegraphics[scale=0.42]
{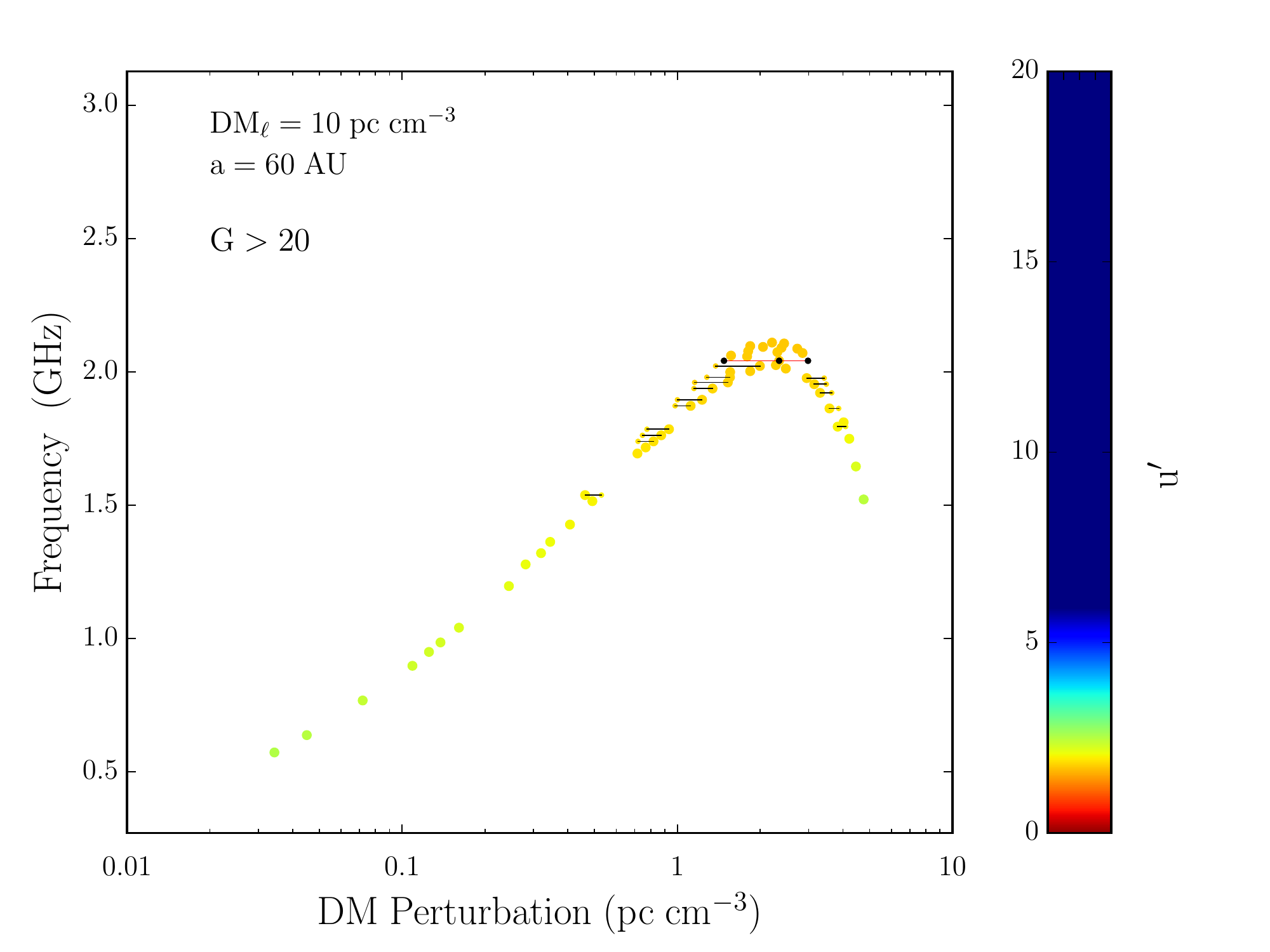}
\caption{Frequencies, arrival times, and dispersion measures of bursts.   Points are shown from locations across the transverse plane $\up$
if they  exceed a minimum gain.
Top row: gains $> 5$.  Bottom row: gains $> 20$.  
The left-hand panels show burst frequencies and arrival times.
The right-hand panels show burst frequencies and dispersion measures. 
The locations of points in the $\up$ plane are color coded according to the color bar.  
Only points with $\up\ge0$ are shown. 
Black  horizontal lines designate `doublets' with the same $\nu$ and $\up$ that both exceed the minimum gain but differ in arrival time and DM. 
Triplets are shown as 
red horizontal lines connecting three black points.  
\label{fig:events_vs_nu_toa_dm}
}
\end{center}
\end{figure*}

To compactly display the range of gains, arrival times, and dispersion measures of bursts,  we show  
events in the frequency-$\delta$TOA plane in Figure~\ref{fig:events_vs_nu_toa_dm}  (left-hand panels) and in the frequency-DM plane (right-hand panels) 
color coded by $\up$.   By symmetry, we  show results only for $\up\ge 0$.
The upper row shows all events with $G>5$ and the lower row with $G>20$.    Many of the events in the upper row are cases with only one image or only one subimage  above the gain threshold.   However, there are many  doublet cases where two events occur at the same frequency and location but arrive at different times.  These are designated by black horizontal lines that connect the pair of events in each case.   There is a smaller number of triplet cases, shown as red horizontal  lines
that connect black points.   Most of the triplets occur near the focal frequency $\nuf \sim 2$~GHz and with arrival times shifted by $\lesssim 10$~ms.   However, doublets occur over a broader range of arrival times up to 40~ms.  

 
Over a range of epochs in which the $\up$ prime plane is sampled,  bright events will occur with arrival times and DMs that are perturbed by the plasma lens.  Bright singlet events will cluster around an arrival time 
$\sim 1$~ms and $\DM\sim 2$~pc~cm$^{-3}$.  Multiplet bursts occuring  at the same epoch will have slightly different 
arrival times and dispersion measures, with spreads up to a few milliseconds and a few $\times$~pc~cm$^{-3}$. 
 
For the particular lens and geometry considered here as an example,  FRBs having  intrinsically  single component    widths $W\sim 1$~to~8 ms will appear as blends of two or three subcomponents when they are imaged as multiplets.   Triplets in some cases may appear as distinct subcomponents with little or no  overlap and slightly different DMs. 


For comparison we also  considered a smaller lens with smaller dispersion depth that has a larger focal frequency. 
This case yields smaller perturbations of arrival times and burst DMs.    Delays are $\lesssim 1~\mu$s at 1~GHz
and differential TOAs between multiply-imaged bursts $\lesssim 0.1~\mu$s can produce frequency structure in the radio 
spectrum on the scale of tens to hundreds of MHz. 


\section{Filaments in the Crab Nebula}
\label{sec:crab}

Here we consider the effects of the  Crab Nebula on measured pulses from the Crab pulsar  because they demonstrate the overall approach taken in this paper. 

\citet[][]{2000ApJ...543..740B} and \citet[][]{2011MNRAS.410..499G} reported transient echoes of radio pulses from the Crab pulsar that they attributed to plasma lensing like that described in  \citet[][]{1998ApJ...496..253C}.  The filaments causing the echoes have estimated  radii $\sim 1$~AU and dispersion depths $\DMlens \sim 0.1$~pc~cm$^{-3}$.   

At 1~GHz the implied lens parameter for a pulsar-lens distance $\dsl = 1$~pc and a pulsar-Earth distance $\dso = 2$~kpc  is 
$\alpha \sim 0.34\, {\dsl}_{\rm pc} \aau^2 (\DMlens / 0.1~{\rm pc~cm^{-3}}) $
corresponding to a focal distance $\df \sim 13$~kpc, well beyond the Crab pulsar's distance.  
However, \citet[][]{2011MNRAS.410..499G} use the duration of the `shadow' region where the flux density is suppressed to estimate the size of the lens.  
Inspection of Figures~\ref{fig:gain} and \ref{fig:gainslices} indicates that the half width of the shadow region is about
twice the lens radius (which is unity in normalized coordinates), so the filaments are a factor of two smaller, leading to a focal distance $\sim 3$~kpc at 1~GHz. 
At the observation frequencies $\nu = 0.33$ and 0.61~GHz where echoes were prominent, the focal distances 
are then 0.35 and 1.2~kpc, consistent with plasma lensing as the cause of the echoes because the Earth is then beyond the focal point and the criterion in Eq.~\ref{eq:at_focal} is met. 

Compared to the filaments causing radio echoes, 
optical filaments in the Crab Nebula have larger  scales extending to 2000~AU \citep[][]{1985ARA&A..23..119D} and potentially large $\DMlens$, giving  
much larger focal distances,  $\df \sim 0.52\,{\rm Gpc}\, (10~{\rm pc~cm^{-3}} / \DMlens) (a / 2000~{\rm AU})^2$. 
This suggests that compact objects in extragalactic supernova remnants may show lensing from similar  structures.  If lensing is required for detection of FRBs,  particular scale sizes and electron densities will be selected as a function of true source distance.   Equivalently, if burst sources reside in supernova remnants of similar age and other properties,  nearby sources will be selected against if they are closer than the focal distance because lensing will not enhance their flux densities. 

\newcommand{\nefour}{n_{\rm e_4}}
\newcommand{\Tfour}{T_{4}}

\section{Plasma Lenses in Host Galaxies}
\label{sec:hosts}

While the IGM provides long path lengths through ionized gas, there is no indication that it harbors small scale structures
 (e.g. 1 to 1000 AU).  However, host galaxies of FRBs almost certainly  have plasma structures similar to those inferred  to be present in the ISM of the Milky Way. We therefore consider   plasma lenses   in host galaxies.  Our treatment is for
 Euclidean space (i.e. low redshifts) because the main points can be made in this regime and the best studied source to date, 
FRB121102,  is at a low redshift $z\sim 0.193$ \citep[][]{2017ApJ...834L...7T}.

Several cases may be considered that represent a large range of possible scale sizes and 
dispersion-measure depths but together can give  similar
values of $\alpha$ (Eq.~\ref{eq:alpha}) and hence the same focal distances and focal frequencies defined earlier. 

One possibility comprises a bow shock produced by  an FRB source moving  in a dense medium; the source-lens distance is then quite small, $\dsl \lesssim 1$~pc.    In this case the FRB source directly influences its plasma environment.   However, the relative velocity between the bow shock and the source is small unless clumps form in the rapidly moving flow along the bow shock.  

A second case is  an FRB source enclosed by a  supernova remnant (SNR) of $\sim$pc radius.  This would satisfy 
Eq.~\ref{eq:df} with small knots or filaments within the shell that provide DM$\sim 1$~\DMunits\ (after accounting for any redshift dependences).    Among SNRs, we expect diverse properties of ionized gas clumps that depend on the masses
of progenitors and on the details of pre-supernova mass loss.

A third possibility is that  larger structures, such as HII regions, with larger $\DMlens$ and larger (e.g. kpc) distances from the source provide the lensing structures.  These would clearly be uninfluenced by the FRB source.   This case may rely on a fortuitious alignment of the lenses with the source while the other options involve direct coupling of the source and the lensing region.    However, FRBs may reside in galaxies where such alignments are not improbable.

To assess the above  three  cases we require that the source and lens be further than the focal distance (i.e. $\df \le \dso$).
  Using Eq.~\ref{eq:df}  and expressing   $\DMlens = a n_{\rm e}$  in terms of the electron density $n_{\rm e} = 10^4\nefour$~cm$^{-3}$, we get 
\be
a \le \frac{\re c^2}{\pi\alphamin} \frac{\dsl n_{\rm e}}{\nu^2} 
= 74~{\rm AU}\times \frac{\nefour \dslkpc}{\nu^2}. 
\label{eq:alimit}
\ee
Lenses at $\sim$kpc distances can be much larger and provide more gain, but still require large electron densities.
However,  lenses within 1~pc of the source must be very small ($\lesssim 0.1$~AU) or the electron density very high ($> 10^4$~cm$^{-3}$) for larger lenses to provide strong lensing effects.    Small lenses provide less gain than larger lenses because the maximum gain $\propto a$ scales as the one-dimensional `area.'

Additional constraints may allow us to favor lenses near or far from a source (but still in the host galaxy).  
Higher gain lenses will show more rapid variability (Eq.~\ref{eq:tcaustic}), all else being equal, because the caustic time scale $\tcaustic \propto G^{-1}$ if the maximum gain $G_{\rm max} \propto a$ is substituted. 

The emission measure,
\be
\EM = a n_{\rm e} = 485~{\rm pc~cm^{-6}} \aau \nefour^2,
\ee
yields  free-free absorption with an optical depth at $\nu$ in GHz \citep[][]{2011piim.book.....D}, 
\be
\tau_{\rm ff} \approx  3.4\times10^{-7} \nu^{-2.1} T_4^{-1.32} \EM. 
\ee
The focusing constraint in 
Eq.~\ref{eq:alimit} implies an upper bound on EM and the optical depth, 
\be
\EM \lesssim 3.59\times 10^4~{\rm pc~cm^{-3} }\times \frac{\nefour^3 \dslkpc}{\nu^2}
\ee
and 
\be
\tau_{\rm ff}  \lesssim 0.12 \times \frac{\nefour^3 \dslkpc}{\nu^2}.
\ee
Free-free absorption can be important  at GHz frequencies for lenses at $\dsl \sim 1$~kpc with densities 
 $\sim 5\times10^4$~cm$^{-3}$ but smaller source-lens distances require significantly higher densities.

The large implied electron density  in all cases gives a large over pressure
${\cal P} = n_{\rm e} kT$ compared to  typical ISM values, identical to the same overpressure identified for Galactic
ESEs \citep[e.g.][]{2016Sci...351..354B}.   The magnetic field that could confine lenses with temperature
$T = 10^4 \Tfour$ is
\be
B = \left(8\pi n_{e} kT \right)^{1/2} \sim 0.59~{\rm mG} \left(\nefour \Tfour\right)^{1/2}
\ee
implying a  maximum Faraday rotation measure (when  the parallel magnetic field is along the line of sight without any reversals),
\be
\RM_{\rm max} \sim  24\,{\rm rad~m^{-2}}\, \aau \nefour^{3/2} \Tfour^{1/2}.
\ee


As yet there is insufficient information to favor lenses for the repeating FRB that are at distances from the source of pc or kpc (or something in between).    Wideband spectra and more complete sampling of periods of high gain can constrain the lens size and dispersion measure depth. 


\section{Summary and Conclusions}
\label{sec:summary}

In this paper we have shown that small, one-dimensional  plasma lenses with small dispersion depths can strongly amplify radio bursts emitted at Gpc distances.   Filamentary structures seen in the Crab Nebula,  timing variations of the Crab pulsar and a few other pulsars, and extreme lensing events in AGN light curves demonstrate that AU-size structures with high electron densities exist
in the Milky Way.   Presumably they also exist in the ionized gas in FRB host galaxies.   
Recent work \citep[][]{2017arXiv170102370M, 2017arXiv170208644B} has made the compelling case that FRBs are associated with young, rapidly spinning magnetars.   These objects dissipate much more energy into their environment than the Crab pulsar and therefore can be expected to drive shocks into presupernova wind material that may fragment into lensing structures.  

In this paper we analyzed the properties of individual plasma lenses with Gaussian density profiles and identified some key features. 
Gaussian plasma lenses  yield both amplification and suppression of the apparent burst flux densities  in
a time sequence as follows.   The gain is initially unity for lines of sight far from the lens. 
As the line of sight nears the lens,  a short duration, large amplitude spike ($\sim 1~$d) occurs followed by   a long-duration ($\sim$months) trough of low gain that can be much less than unity.   Another caustic spike is then encountered that asymptotes to the quiescent state as the line of sight moves away from the lens.    These time scales can be larger or smaller depending on the size of the lens and on the effective velocity by which the geometry changes. 

As a function of frequency,  the gain pattern suppresses all frequencies in the trough but shows asymmetric, large amplitude cusps  at epochs where a source is multiply imaged.  Burst spectra can therefore show narrow structure even if  the intrinsic spectrum is smooth.

 Gaussian density profiles produce multiply imaged  bursts with different apparent strengths,   arrival times, and dispersion measures.      The frequency dependence of arrival times can differ markedly from the 
$\nu^{-2}$ scaling of cold plasma dispersion.    If arrival times of individual bursts are fitted  with a $\nu^{-2}$ scaling law, negative dispersion measure perturbations will be obtained at some epochs (which, of course, add to a much larger DM through the remainder of the host galaxy,  the IGM, and the Galaxy).   When subimages are comparable in strength, 
burst spectra may also show  oscillations due to interference between subimages if their arrival time difference is smaller than the burst width.  Arrival time perturbations from lenses, whether for a singly imaged  or multiply imaged burst, can be large enough to mask any  underlying periodicity with period much less than one second. 

Detection of bursts in large-scale surveys may rely on the presence of spectral caustics, which can elevate the apparent flux density by factors up to $\sim 100$.  Followup observations, in turn, may be deleteriously affected by the gain trough that will follow a caustic spike.  The actual amount of amplification is larger for larger lenses and will be attenuated for sources larger than about $10^{10}$~cm for nominal parameters.   Any scattering from additional electron density fluctuations in the lens or between the source and lens  can also attenuate the amplification.   

Many of the phenomena observed from  the repeating FRB121102 are consistent, at least qualitatively,  with the features we have derived for the Gaussian lens. These, in turn,    are expected to be generic for lenses with non-Gaussian column density profiles because they result primarily from the presence of inflection points in the total phase (geometric + dispersive).  
Non-Gaussian lenses will share some of these properties but likely will not be symmetric nor do we expect the specific details of  cusps in the spectra or light curves of radio sources to be the same.     

Testing of the lensing interpretation of the intermittency of FRB121102 (and other FRB sources if they actually repeat) can proceed in several ways, including the measurement of very wideband spectra covering  many octaves, e.g. 0.4 to 20 GHz. 
These can identify the pairs of caustic spikes expected from a single Gaussian lens or, perhaps, multiple spikes for more complicated lensing structures.     If such spectra are obtained with high spectral resolution ($< 1$~MHz), oscillations from interference between subimages may also be detected in frequency ranges containing caustic spikes. 

A detailed discussion of individual bursts from FRB121102, including an interpretation in terms of lensing, will be given in papers by J. Hessels et al.  (in preparation) and C. Law et al. (in preparation).    Further study of lensing that include two dimensional lenses and the effects of multiple lenses is also in progress. 

A related analysis should be made of Galactic lenses to address why only modest  lensing gains  have been  identified (so far)  from Galactic sources.   It is possible that most Galactic  lenses have very small or very large focal distances that are selected against.    Radio wave scattering from the general ISM may also mask or attenuate strong lensing. 

Acknowledgements: 

We thank Laura Spitler for useful correspondence and conversations. 
S.C. and J.M.C.,  are partially supported by the NANOGrav Physics Frontiers Center (NSF award 1430284). Work at Cornell (J.M.C., S.C., R.S.R. ) was supported by NSF grants AST-1104617 and AST-1008213.
J.W.T.H. acknowledges funding from an NWO Vidi fellowship and from
  the European Research Council under the European Union's Seventh
  Framework Programme (FP/2007-2013) / ERC Starting Grant agreement
  nr. 337062 ("DRAGNET"). 
 I. W. was supported by National Science Foundation Grant Phys 1104617 and National Aeronautics and Space Administration grant NNX13AH42G.
  Part of this research was carried out at the Jet Propulsion Laboratory, California Institute of Technology, under a contract with the National Aeronautics and Space Administration. 








\begin{appendix}
\section{Kirchhoff Diffraction Integral (KDI) for the Gaussian Plasma Lens}
\label{app:KDI}

\newcommand{\Gpthree}{G_{\rm p_3}}


We use  dimensionless coordinates $u$, $\us$, $\up$ defined in the main text  and 
 $\uF$,   the normalized Fresnel scale ($\rF$),
\be
\uF = \frac{\rF}{a} = 
	\frac{1}{a} \left(\frac{\lambda \dsl \dlo}{2\pi\dso}\right)^{1/2}.
\label{eq:uF}
\ee
The scalar field from the KDI is
\be
\varepsilon(\up) &=& \frac{1}{\uF \sqrt{2\pi i} }  \int du\, \exp \left [  {i \Phi(u, \us, \uobs)} \right],
\ee
where the total phase is
\be
\Phi(u, \us, \uobs) &=& \left( 2\uF^2 \right)^{-1} 
		\left [ 
		(\dlo/\dso) (u-\us)^2 + (\dsl/\dso) (u-\uobs)^2 - (\dsl\dlo/\dso^2)(\uobs-\us)^2 
		\right]
		+ \phi_{\rm lens}(u)
\ee
and  $\phi_{\rm lens}(u)$ is the phase perturbation from the lens. 
The KDI is normalized so that without  a screen ($\phi_{\rm lens}=0$), the intensity $\vert \varepsilon(\up)  \vert^2 = 1$.
The lens equation in Eq.~\ref{eq:lens_eq} (main text) is  the stationary phase (SP) solution given by $\partial_u \Phi(u, \us, \uobs) = 0$
where $\partial_u \equiv \partial/\partial u$.

The geometrical optics gain $\Gg$  (Eq.~\ref{eq:gain})  results from expanding  the phase to second order  about a particular SP solution $\ubar$,  $\Phi(u) \approx \Phi(\ubar, \us, \uobs) + (1/2) \partial_u^2 \Phi(\ubar, \us, \uobs) (u-\ubar)^2$  and taking the squared magnitude of  the resulting integral.   We  define $\Gg$ to be $\ge 0$  but also define the parity of the integral as the sign of the second derivative, which changes at the inflection point of the total phase.  
 
 When the geometrical optics gain diverges, higher order terms in the phase expansion (or the exact phase)  are needed to properly calculate the gain. Numerical and analytical results indicate that for $\Gg\to\infty$,  the physical optics gain is  given  for most $\ubar$ by using  the additional cubic  term $(1/6)\partial_u^3 \Phi(\ubar, \us, \up) (u-\ubar)^3$ in the KDI, 
which yields the analytical result, 
 \be
 \Gpthree \approx \frac{6^{2/3}2}{9\pi} 
 	\left[
	    \frac{\Gamma(1/3) \cos \pi/6}{\vert\partial_u^3 \Phi(\ubar, \us, \uobs)\vert^{1/3}}
	\right]^2.
\label{eq:G3}
 \ee
 If $\partial_u^3 \Phi$ has any zeros in $u$, higher-order terms are needed. 

For the Gaussian lens,  
$\phi_{\rm lens}(u) =  - \lambda \re \DMlens e^{-u^2} = -(\alpha/ 2\uF^2)  e^{-u^2}$.
The geometrical optics gain is 
\be
\Gg = \vert 1 + \alpha (1-2u^2) e^{-u^2} \vert^{-1}.
\ee
At  inflection points where $\Gg \to \infty$, $\Gpthree$ is calculated instead by
substituting the third derivative $\partial_u^3 \Phi(\ubar, \us, \uobs) = 4\uF^{-2} \alpha \ubar e^{-\ubar^2}(\ubar^2 - 3/2)$ into Eq.~\ref{eq:G3}. The approximation is good except at or near $\ubar=\sqrt{3/2}$ where $\Gpthree$  diverges.  In that
case we evaluate the KDI to obtain a finite value for the physical optics gain,  
$\Gp(\ubar=\sqrt{3/2}) \sim \uF^{-1} \propto a$ for $\alpha_\infty = e^{3/2} / 2 = 2.24$. 
We note that for this case where $\ubar=\sqrt{3/2}$, the KDI is oscillatory because this location with  $\alpha = \alpha_\infty$ produces an  optical catastrophe \citep[e.g.][]{Berry1980257}. 


\end{appendix}


\end{document}